\documentclass[aps,prb,preprint,showpacs,byrevtex]{revtex4}
\let\oldhat\hat
\renewcommand{\hat}[1]{\oldhat{\mathbf{#1}}}

\usepackage{times,mathptm}
\usepackage[dvips]{graphicx,color}

\begin{document}
\title{Giant spin-orbit-induced spin splitting in Bi zigzag chains on GaAs(110)}
\author{Hyun-Jung Kim$^{1}$ and Jun-Hyung Cho$^{1,2*}$}
\affiliation{$^1$ Department of Physics and Research Institute for Natural Sciences, Hanyang University,
17 Haengdang-Dong, Seongdong-Ku, Seoul 133-791, Korea \\
$^2$ International Center for Quantum Design of Functional Materials (ICQD), HFNL,
University of Science and Technology of China, Hefei, Anhui 230026, China}
\date{\today}

\begin{abstract}
The search for one-dimensional electron systems with a giant Rashba-type spin splitting is of importance for the application of spin transport. Here we report, based on a first-principles density-functional theory calculation, that Bi zigzag chains formed on a heterogeneous GaAs(110) surface have a giant spin splitting of surface states. This giant spin splitting is revealed to originate from spin-orbit-coupling (SOC) and electric dipole interaction that are significantly enhanced by (i) the asymmetric surface charge distribution due to the strong SOC-induced hybridization of the Bi $p_x$, $p_y$, and $p_z$ orbitals and (ii) the large out-of-plane and in-plane potential gradients generated by two geometrically and electronically inequivalent Bi atoms bonding to Ga and As atoms. The results demonstrate an important implication of the in-plane and out-of-plane asymmetry of the Bi/GaAs(110) interface system in producing the giant spin splitting with the in-plane and out-of-plane spin components.
\end{abstract}

\maketitle

\vspace{0.4cm}
\section{INTRODUCTION}
\vspace{0.4cm}

Spintronics is one of the most growing research fields in condensed matter physics~\cite{jansen}. Recently, the Rashba-type spin splitting~\cite{bychkov}, where the spin-orbit coupling (SOC) lifts the spin degeneracy in the inversion-symmetry broken environments such as surfaces of solids or interfaces of heterostructures, has drawn much attention because of the possibility of coherent spin manipulation without external magnetic field. After the first observation of such spin-split bands at the Au(111) surface~\cite{lashell}, a number of experimental and theoretical studies have intensively examined the Rashba-type spin splitting at various heavy-metal surfaces~\cite{koroteev}, heavy-metal overlayers on surfaces~\cite{ast,gierz,varykhalov,barke,tegenkamp,park}, and semiconductor heterostructures~\cite{nitta}. Especially, the search for one-dimensional (1D) spin-polarized electronic systems is desirable for manipulating spin carriers~\cite{sasaki,jansen}. In this regard, 1D nanowires formed on Si surfaces such as Au chains on stepped Si surfaces~\cite{barke}, Pt nanowires on Si(110)~\cite{park}, and Pb atom wires on Si(557)~\cite{tegenkamp} have been investigated to show a relatively larger Rashba-type spin splitting compared to two-dimensional (2D) electron systems.

Although the Rashba-Bychkov model~\cite{bychkov} has been remarkably successful in explaining the phenomena of the Rashba spin splitting in terms of the effective magnetic filed experienced by electrons moving in the potential gradient perpendicular to the surface, it significantly underestimates the spin splitting as an energy scale of ${\sim}$10$^{-6}$ eV which is about 10$^5$ times smaller~\cite{petersen} than experimental measurements~\cite{lashell}. In order to estimate quantitatively the size of the Rashba-type spin splitting, several recent studies have taken into account other factors missed in the Rashba-Bychkov model, such as the atomic SOC effect~\cite{petersen}, the in-plane anisotropy of surface potential~\cite{ast,gierz,premper}, the asymmetry of surface charge distribution~\cite{bihlmayer2,nicolay,nagano,krupin}, and the presence of orbital angular momentum at surfaces~\cite{park1,park2,park3}. Since such various contributions to the Rashba-type spin splitting can be enhanced in 1D electron systems compared to 2D electron systems~\cite{park,takayama}, it is interesting and challenging to search for the 1D electron system with a giant Rashba-type spin splitting. Here, by choosing a prototypical 1D electron system formed on a heterogeneous semiconductor substrate, we propose an ideal platform for exploration of the giant spin splitting.

\begin{figure}[ht]
\includegraphics[width=0.48\textwidth]{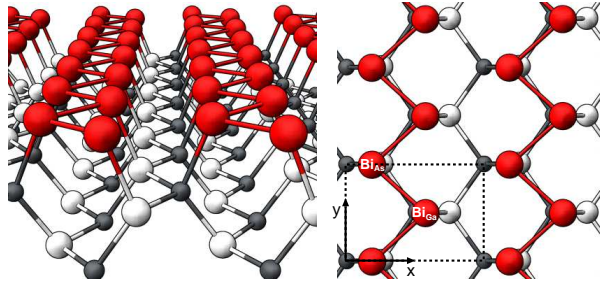}
\caption{(Color online) Perspective and top views of the optimized structure of Bi zigzag chains on a heterogeneous GaAs(110) surface. The dotted line indicates the unit cell. The $x$ and $y$ axes point along the [001] and [1$\overline{1}$0] directions, respectively. The large, medium, and small circles represent Bi, Ga, and As atoms, respectively. Two Bi atoms bonding to Ga and As atoms are labeled as Bi$_{\rm Ga}$ and Bi$_{\rm As}$, respectively.}
\end{figure}

In this paper, we present a density-functional theory (DFT) study of the giant spin splitting in self-assembled Bi zigzag chains on GaAs(110). This zigzag chain is composed of two geometrically and electronically inequivalent Bi atoms bonding to Ga and As atoms (see Fig. 1)~\cite{mclean,joyce}, forming an interface at which the broken inversion symmetry produces the large potential gradients perpendicular and parallel to the surface. We find that the SOC enhances the hybridization of the Bi $p_x$, $p_y$, and $p_z$ orbitals as well as Ga and As $p$ orbitals, giving rise to the asymmetry of surface charge distribution around the Bi atoms and the Ga and As substrate atoms up to the third deeper substrate layer. This significant asymmetric surface charge distribution together with the out-of-plane and in-plane surface electric fields increases the magnitudes of SOC and electric dipole interaction, leading to a giant spin splitting with the in-plane and out-of-plane spin components. The present findings will provide important implications for understanding the underlying driving forces behind the possible giant spin splitting in 1D nanowires formed on heterogeneous semiconductor substrates.

\vspace{0.4cm}
\section{CALCULATIONAL METHOD}
\vspace{0.4cm}

The present DFT calculations were performed using the Vienna {\it ab initio} simulation package with the projector-augmented wave method and a plane wave basis set~\cite{vasp1,vasp2}. For the treatment of exchange-correlation energy, we employed the generalized-gradient approximation functional of Perdew-Burke-Ernzerhof (PBE)~\cite{pbe}. The Bi/GaAs(110) substrate was modeled by a periodic slab geometry consisting of the seven Ga and As atomic layers with ${\sim}$28 {\AA} of vacuum in between the slabs. The bottom of the GaAs substrate was passivated by pseudohydrogen atoms~\cite{shiraishi} with 0.75 or 1.25 $e$. We employed a dipole correction that cancels the artificial electric field across the slab~\cite{neugebauer}. The ${\bf k}$-space integration was done with the 24${\times}$36 Monkhorst-Pack meshes in the surface Brillouin zones (SBZ) of the $1\times1$ unit cell. All atoms except the bottom two substrate layers were allowed to relax along the calculated forces until all the residual force components were less than 0.02 eV/{\AA}.

\vspace{0.4cm}
\section{RESULTS}
\vspace{0.4cm}

We begin to optimize the atomic structure of the Bi/GaAs(110) surface using DFT calculation within the generalized gradient approximation of Perdew-Burke-Ernzerhof (PBE) in the absence of SOC. The optimized structure is displayed in Fig. 1. The calculated bond lengths at the interface are $d_{\rm Bi-Bi}$ = 3.00 {\AA}, $d_{\rm Bi-Ga}$ = 2.77 {\AA}, and $d_{\rm Bi-As}$ = 2.77 {\AA}, in good agreement with previous theoretical~\cite{umerski2} and experimental data~\cite{ford} (see Table IS of the Supplemental Material~\cite{supple}). It is noticeable that two Bi atoms bonding to Ga and As atoms (designated as Bi$_{\rm Ga}$ and Bi$_{\rm As}$ in Fig. 1) are geometrically and electronically inequivalent because the Bi$-$Ga and Bi$-$As bonds have some different ionic characters due to a larger electronegativity of As compared to Ga and Bi atoms. Here, Bi$_{\rm Ga}$ positions 0.11 {\AA}-higher than Bi$_{\rm As}$, and it has a 5$d$ core-level shift of 0.31 eV to lower binding energy relative to that of Bi$_{\rm As}$, in accordance with the measured value of ${\sim}$0.4 eV from x-ray photoemission spectroscopy~\cite{joyce,schaffler}. Due to these asymmetric characters of the Bi chains, the Bi/GaAs(110) interface produces the out-of-plane and in-plane potential gradients, as discussed below. Figure 2(a) shows the calculated band structure of Bi/GaAs(110) with the band projection onto the $p_x$, $p_y$, and $p_z$ orbitals of Bi. It is seen that the band dispersion of the highest occupied surface state (hereafter, designated as $SS$) is nearly flat along the $\overline{{\Gamma}X}$ and $\overline{YM}$ lines while broad along the $\overline{{\Gamma}Y}$ and $\overline{XM}$ lines, indicating a 1D electronic state~\cite{hu1,hu2} along the Bi chains. As shown in Fig. 2(a) and 2(b), the orbital character of the $SS$ state exhibits a strong ${\bf k}$-dependence within the surface Brillouin zone (SBZ): e.g., the $p_x$ or $p_z$ character along the $\overline{{\Gamma}Y}$ line, $p_x$ along the $\overline{{\Gamma}X}$ line, $p_x$, $p_y$, or $p_z$ along the $\overline{XM}$ line, and $p_z$ along the $\overline{MY}$ line. As a consequence of such complex ${\bf k}$-dependent orbital characters of the $SS$ state, SOC easily induces a hybridization of the Bi $p_x$, $p_y$, and $p_z$ orbitals, as demonstrated below.

\begin{figure}[ht]
\includegraphics[width=0.44\textwidth]{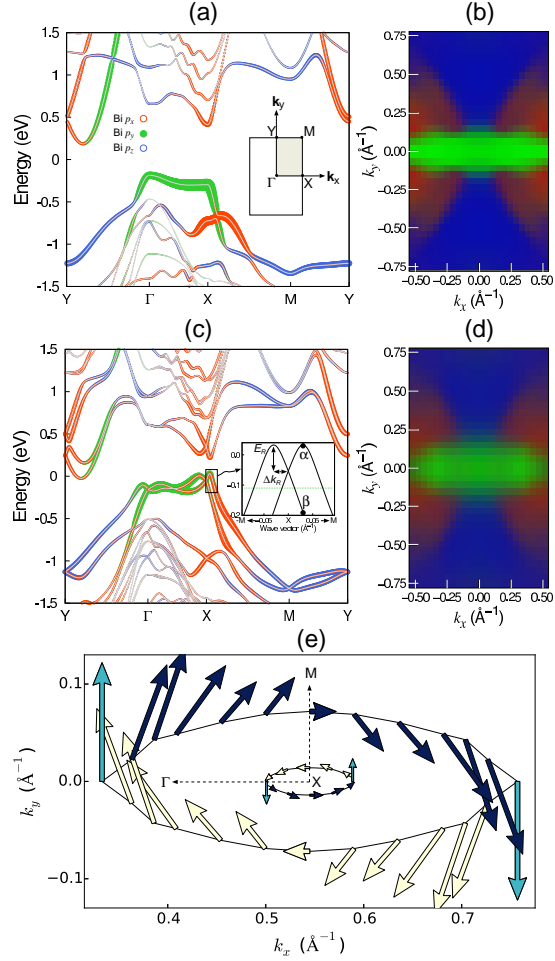}
\caption{(Color online) (a) Band structure and (b) Bi $p_x$, $p_y$, and $p_z$ orbital characters of the $SS$ state obtained using PBE. The corresponding results obtained using PBE+SOC are given in (c) and (d). In (a) and (c), the bands projected onto Bi $p_x$, $p_y$, and $p_z$ orbitals are displayed. Here, the radii of circle are proportional to the weights of corresponding orbitals. The energy zero represents the Fermi level $E_F$. The inset in (a) shows the SBZ of the unit cell, while that in (c) magnifies the energy dispersion of the $SS$ state along the $X{\rightarrow}M$ and $X{\rightarrow}$$-M$ directions. In (b) and (d), the $p_x$, $p_y$, and $p_z$ orbital components are mapped by using red, green, and blue color channels with their brightness, respectively. In (e), constant-energy contours around the $X$ point, taken at 0.11 eV below $E_F$, are plotted with spin texture. Here, the arrows with increasing their brightness represent the SAMs with the negative, zero, and positive out-of-plane components, respectively.}
\end{figure}

Next, we examine the effect of SOC on the geometry and band structure of Bi/GaAs(110) using the PBE+SOC calculation. It is found that the inclusion of SOC changes $d_{\rm Bi-Bi}$, $d_{\rm Bi-Ga}$, and $d_{\rm Bi-As}$ by less than 0.04 {\AA} (see Table IS of the Supplemental Material~\cite{supple}). Figure 2(c) shows the PBE+SOC band structure of Bi/GaAs(110)~\cite{hseresult}. The spin degeneracy of the $SS$ state as well as other states is found to be lifted over the SBZ except at the high-symmetry points (i.e., ${\Gamma}$, $X$, $M$, and $Y$ points). Obviously, the PBE+SOC band projection [Fig. 2(c)] and orbital character [Fig. 2(d)] of the $SS$ state indicate a strong hybridization between the Bi $p_x$, $p_y$, and $p_z$ orbitals, compared to the PBE cases [see Fig. 2(a) and 2(b)], leading to a electric dipole interaction that will be discussed in detail later. The inset of Fig. 2(c) shows a close up of the spin splitting near the $X$ point along the $\overline{XM}$ direction, illustrating the characteristic dispersion of a Rashba-type spin splitting. For comparison of the size of the spin splitting with those of previously reported Rashba systems~\cite{park,barke,lashell,koroteev,ishizaka,takayama,ast}, we fit the $k$-dependent dispersion of the spin-split subbands of the SS state along the $k_x$ or $k_y$ direction by using the Rashba Hamiltonian $H$ = $\frac{{\bf p}^2}{2m^*}$ + $\frac{{\alpha}_R}{\hbar}({\bf p}{\times}{\hat z}){\cdot}{\bf S}$, where $m^*$ is the electron effective mass, ${\alpha}_R$ the Rashba parameter, and ${\bf p}$ and \textbf{\emph {S}} are a momentum operator and a spin angular momentum (SAM) operator, respectively. Here, the spin-dependent eigenvalues become ${\epsilon}_{\pm}$ = $\frac{{{\hbar}^2}k^2}{2m^*} {\pm} {{\alpha}_R}k$ with the different values of $m^*$ and ${\alpha}_R$ along the the $k_x$ and $k_y$ directions. From the calculated band structure, the characteristic parameters of spin-split subbands such as the momentum offset ${\Delta}k_R$ and the Rashba energy $E_R$ [see the inset of Fig. 2(c)] are extracted to estimate $m^*$ and ${\alpha}_R$ using the Rashba Hamiltonian. The present values of ${\Delta}k_R$, $E_R$, and ${\alpha}_R$ at the high-symmetry points are listed in Table I, together with those of previous Rashba systems~\cite{park,barke,lashell,koroteev,ishizaka,takayama,ast}. We find that ${\alpha}_R$ = 4.94 (2.27) eV {\AA} along $X{\rightarrow}M$ ($\Gamma{\rightarrow}Y$) is larger than 1.86 (1.09) eV {\AA} along $X{\rightarrow}\Gamma$ ($\Gamma{\rightarrow}X$), indicating an enhanced spin splitting along the $y$ direction parallel to Bi chains~\cite{hseresult}. Remarkably, the present values of ${\alpha}_R$ are among the largest ones so far reported for Rashba spin-split systems (see Table I). We note that the constant energy contours around the $X$ point, taken at an energy of 0.11 eV below $E_F$, exhibit a pronounced anisotropy between the $\overline{X{\Gamma}}$ and $\overline{XM}$ directions, yielding concentric ellipses [see Fig. 2(e)]. Moreover, the constant energy contours around the ${\Gamma}$ point at energies lower than ${\sim}$0.2 eV below $E_F$ are open in the $\overline{{\Gamma}X}$ direction (see Fig. 1S of the Supplemental Material~\cite{supple}) because the $SS$ state has a flat band character along the $\overline{{\Gamma}X}$ line [see Fig. 2(c)].

\begin{table}[ht]
\caption{
Calculated Rashba parameters characterizing the spin splitting of the $SS$ state around high symmetry points of the SBZ, together with
those of previously reported Rashba systems. The corresponding values obtained using HSE+SOC~\cite{hseresult} are also given in parentheses.}
\begin{ruledtabular}
\begin{tabular}{lccc}
   		& ${\Delta}k_R$ ($\AA^{-1}$) 	& $E_R$ (meV) 	&   ${\alpha}_R$ (eV {\AA})   	\\  \hline
Bi/GaAs(110)$-$This \\
   \- \ \- \ $\Gamma{\rightarrow}X$ 	&  0.11(0.13) & 62(53)	& 1.09(0.82)  \\
   \- \ \- \ $\Gamma{\rightarrow}Y$ 	&  0.05(0.04) & 51(39)	& 2.27(1.91)  \\
   \- \ \- \ $Y{\rightarrow}\Gamma$ 	&  0.13(0.14) & 71(85)	& 1.09(1.20)  \\
      \- \ \- \ $Y{\rightarrow}M$ 	& 0.06(0.06)  & 10(11)   & 0.33(0.37)   \\
   \- \ \- \ $M{\rightarrow}Y$ 	&  0.05(0.06) & 24(23)	& 0.87(0.82)  \\
   \- \ \- \ $M{\rightarrow}X$ 	&  0.06(0.08) & 19(25)	& 0.66(0.63)  \\
   \- \ \- \ $X{\rightarrow}M$ 	&  0.03(0.03) & 80(75)	& 4.94(5.00)  \\
   \- \ \- \ $X{\rightarrow}\Gamma$  & 0.06(0.06) & 60(31)	& 1.86(1.09)  \\
Au/Si(557) (1D)~\cite{barke} & 		0.05			&  $-$		&  $-$					\\
Pt/Si(110) (1D)~\cite{park}	& 		0.12				& 81			& 1.36					\\
Bi/Si(111) (1D)~\cite{takayama}	& 		0.17				& 68			& 0.80					\\
Au(111) (2D)~\cite{lashell} & 		0.012			& 2.1			& 0.33 					\\
Bi(111) (2D)~\cite{koroteev} & 		0.05		& 14			& 0.55 					\\
Bi/Ag(111) (2D)~\cite{ast} & 		0.13		& 200			& 3.05 					\\
BiTeI (3D)~\cite{ishizaka}  & 		0.052			& 100			&  3.8 					\\
\end{tabular}
\end{ruledtabular}
\end{table}

On the experimental side, angle-resolved photoemission spectroscopy~\cite{mclean} observed a nearly flat surface state along the $\overline{{\Gamma}X}$ line and the other surface state between the midpoint of the $\overline{{\Gamma}X}$ line and the X point. The former surface state is ${\sim}$0.8 eV above the latter one. These dispersion features of the two surface states are generally similar to our band structure obtained using the PBE or PBE+SOC calculation. In addition, according to the inverse photoelectron spectroscopy (IPS) experiment of McLean and Himpsel~\cite{mclean2}, the Bi monolayer on GaAs(110) produced two unoccupied surface states around 0.9 and 1.9 eV above the bulk valence-band maximum (VBM) at the ${\Gamma}$ point. On the other hand, the IPS experiment of Hu $et$ $al$.~\cite{hu2} observed a pronounced Bi-derived surface resonance state around 1.25 eV above $E_F$ at the ${\Gamma}$ point. Noting that the band gap of the GaAs bulk is about 0.9 eV, the 1.25 eV-peak in the latter IPS data~\cite{hu2} may correspond to the second peak in the former IPS data.~\cite{mclean2} From our PBE (PBE+SOC) band structure, it is found that the positions of two lower unoccupied states at the ${\Gamma}$ point are 1.12 (1.12) and 1.56 (1.45) eV above the bulk VBM, respectively, in reasonable agreement with previous IPS experiments~\cite{mclean2,hu2}.

\begin{figure}[t]
\includegraphics[width=0.45\textwidth,angle=0]{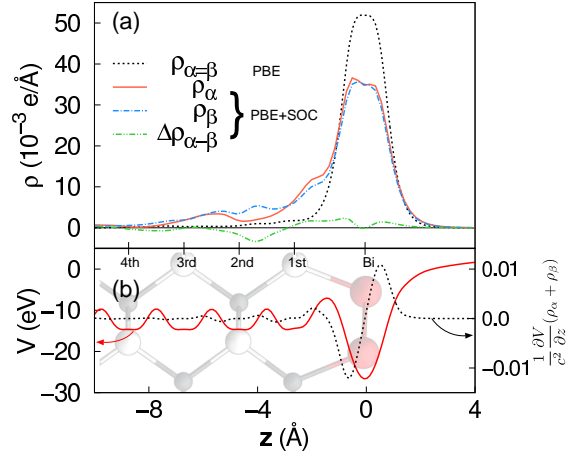}
\caption{(Color online) (a) Planar-averaged electron charge densities for the $SS$ state at $k_0$ [see the inset of Fig. 2(c)], obtained using PBE and PBE+SOC. In (b), the planar-averaged crystal potential and $\frac{1}{c^2} \frac{{\partial}V}{{\partial}z} {({\rho}_{\alpha}+{\rho}_{\beta})}$ (in Ryd atomic units) at $k_0$ are displayed along the $z$ direction.}
\end{figure}

Recently, it has become known that the asymmetric features of the surface states at surface atoms are crucial to determine the size of a Rashba spin splitting through SOC and electric dipole interaction. The former Hamiltonian is given by $H_{\rm SOC}$ = $\frac{2}{c^2} ({\nabla}V{\times}{\bf p})\cdot\textbf{\emph {S}}$, where $c$ is the velocity of light and $V$ is a crystal potential, whereas the latter one represents the electrostatic energy of electric dipole moment in the surface electric field. In Fig. 3(a), we plot the $xy$ planar-averaged electron charge densities (${\rho}_{\alpha}$ and ${\rho}_{\beta}$) of the spin-split $SS$ state at $k_0$ = 0.03 {\AA}$^{-1}$ [see the inset of Fig. 2(c)] away from the $X$ point along the $\overline{XM}$ line. It is seen that ${\rho}_{\alpha}$ and ${\rho}_{\beta}$ exhibit some delocalization up to the third deeper GaAs substrate layer. This delocalization feature strikingly contrasts with the spin-unpolarized case of ${\rho}_{{\alpha}={\beta}}$ [see Fig. 3(a)] obtained using the PBE calculation without SOC, which shows a highly localized charge character around Bi atoms. Also we note that the PBE+SOC result for ${\rho}_{\alpha}$ and ${\rho}_{\beta}$ shows the broad maximum consisting of two peaks, possibly due to Bi$_{\rm Ga}$ and Bi$_{\rm As}$ atoms. Thus, we can say that the SOC induces a large asymmetry of surface charge distribution, which in turn contributes to the Rashba spin splitting by ${\Delta}E_{\rm SOC}$ = $\sqrt{{k_x}^2 + {k_y}^2}$ $\int {\rm d}z \frac{1}{c^2} \frac{{\partial}V}{{\partial}z} {({\rho}_{\alpha}+{\rho}_{\beta})}$~\cite{nagano}. Figure 3(b) displays not only the the $xy$ planar-averaged crystal potential $V$ as a function of $z$ but also the calculated integrand of ${\Delta}E_{\rm SOC}$ for the spin-split eigenstates at $k_0$. It is found that the magnitude of ${\Delta}E_{\rm SOC}$ can be dominant at the positions near Bi layer and the first GaAs substrate layer. On the other hand, the contribution of electric dipole interaction to the Rashba spin splitting can be expressed as ${\Delta}E_{\rm D}$ = $-{\Delta}{\bf d}\cdot{\bf E_s}$, where ${\bf E_s}$ denotes the surface electric field and ${\Delta}{\bf d}$ is the dipole moment difference obtained from ${\rho}_{\alpha}$ and ${\rho}_{\beta}$. For ${\rho}_{\alpha}$ and ${\rho}_{\beta}$ at $k_0$, we obtain their difference ${\Delta}{\rho}_{{\alpha}-{\beta}}$ [see Fig. 3(a)] and then calculate ${\Delta}{\bf d}$ = $-$0.19 $e${\AA} along the $z$ direction. Since $V$ around the Bi layer represents a highly deep, asymmetric quantum well [see Fig. 3(b)], a sizable magnitude of ${\Delta}E_{\rm D}$ can be expected by a very large surface electric field. It is noted that the asymmetric features of the heterogeneous GaAs substrate as well as the Bi chains composed of two geometrically and electronically inequivalent Bi$_{\rm Ga}$ and Bi$_{\rm As}$ atoms produce an asymmetric in-plane surface charge distribution as well as an in-plane potential gradient along the $x$ direction (see Fig. 2S of the Supplemental Material~\cite{supple}), thereby contributing to the spin splitting through ${\Delta}E_{\rm SOC}$ and ${\Delta}E_{\rm D}$. Thus, we can say that the SOC-induced asymmetries of the out-of-plane and in-plane surface charge distributions together with the out-of-plane and in-plane surface electric fields lead to the giant spin splitting in the Bi/GaAs(110) surface system.

Figure 2(e) shows the helical spin texture with the in-pane and out-of-plane spin components along the constant-energy contours around the $X$ point, where the SAM direction rotates anti-clockwise (clockwise) at the inner (outer) contour. Note that the spins along the outer contour, located off from the $\overline{X{\Gamma}}$ or $\overline{XM}$ line, have the radial in-plane component (directing perpendicular to constant energy contour line), indicating that the spin splittings involve some contribution of the Dresselhaus effect~\cite{dresselhaus,ganichev}. Such Dresselhaus spin-orbit splittings may arise from the asymmetry of electrostatic potential or charge density in the Bi layer (composed of Bi$_{\rm Ga}$ and Bi$_{\rm As}$) and the heterogeneous GaAs substrate. In addition, the Bi/GaAs(110) system has one mirror-plane ${\sigma}_v$ symmetry with the $xz$ plane (see Fig. 1), which is a combination of the proper rotation of 180$^\circ$ (about the $y$ axis) with the inversion. Therefore, when the spin vectors move along the constant-energy contours from the irreducible part [i.e., the darkened area in the inset of Fig. 2(a)] to the neighboring one through the mirror-plane ${\sigma}_v$ symmetry, the $S_x$ and $S_z$ components change their sign but the $S_y$ component remains unchanged. Consequently, the spin vectors at the points crossing the $\overline{{\Gamma}X}$ line are oriented perpendicular to the mirror plane, and the whole spin texture along the constant-energy contours also satisfies the time-reversal symmetry that reverses simultaneously the wavevector and spin [see Fig. 2(e)].

\begin{figure}[ht]
\includegraphics[width=0.45\textwidth,angle=0]{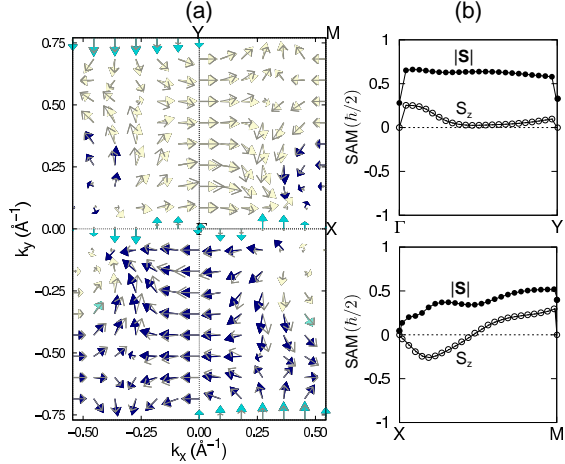}
\caption{(Color online) (a) SAM and OAM textures of the upper subband of the $SS$ state and (b) the total spin polarization $|{\bf \it S}|$ = $\sqrt{{S_x}^2 + {S_y}^2 + {S_z}^2}$ and the $S_z$ component along the $\overline{{\Gamma}Y}$ and $\overline{XM}$ lines. In (a), SAM is drawn with a larger arrow head than OAM. Here, the SAM vectors with the negative, zero, and positive out-of-plane components are represented with increasing brightness of the arrow head. The arrows for OAM represent only the in-plane components. The values in (b) are given in units of ${\hbar}$/2.}
\end{figure}

It is noteworthy that the SOC-induced asymmetric surface charge distributions of the $SS$ state originate from the admixture of the Bi $p_x$, $p_y$, and $p_z$ orbitals [see Fig. 2(c) and 2(d)] as well as their hybridization with Ga and As $p$ orbitals (see Fig. 3S of the Supplemental Material). These orbital mixings in the $SS$ state result in the formation of orbital angular momentum (OAM). Recently, Park $et$ $al$.~\cite{park3} pointed out that the existence of OAM on the surfaces of high atomic number materials produces the electric dipole moment that interacts with the surface electric field, giving rise to a Rashba-type spin splitting. Indeed, as shown in Fig. 4(a), the OAM is present in the upper subband of the $SS$ state and is oriented mostly parallel to the corresponding SAM~\cite{lock}, manifesting the interplay between orbital ordering and SOC for the Rashba-type spin splitting. It is interesting to note that, around the ${\Gamma}$ point and near the $\overline{{\Gamma}Y}$ line, the planar component of spin vectors is mostly composed of $S_x$. This feature of the spin texture is likely to be associated with the band structure of the $SS$ state showing 1D character along the $y$ direction: i.e., the direction of SAM (or OAM) is perpendicular to the electron group velocity ${\bf v}$ = $\frac{1}{\hbar}$${\nabla}_{\bf k}E({\bf k})$ which nearly points in the $y$ direction (see Fig. 4S of the Supplemental Material). Such a locking of SAM (or OAM) to the crystal momentum that maximizes the magnitude of ${\Delta}E_{\rm D}$~\cite{park3} can be utilized for the non-vanishing spin transport along the Bi chains. Figure 4(b) shows the values of the total spin polarization $|{\bf \it S}|$ and the $S_z$ component along the $\overline{{\Gamma}Y}$ and $\overline{XM}$ lines. We find that along the $\overline{XM}$ line the magnitude of $S_z$ is comparable to those of the parallel components. This sizable $S_z$ (or $L_z$) component reflects the presence of the in-plane dipole moment and the in-plane potential gradient (see Fig 2S of the Supplemental Material~\cite{supple}) generated by two electronically different Bi atoms bonding to Ga and As atoms.

\vspace{0.4cm}
\section{SUMMARY}
\vspace{0.4cm}

We have predicted for the first time that Bi zigzag chains self-assembled on a heterogeneous GaAs(110) surface have a giant spin splitting with the in-plane and out-of-plane spin components. By means of the DFT calculations, we revealed that this giant spin splitting originates from SOC and electric dipole interaction which are significantly enhanced by the large asymmetric surface charge distribution and the large out-of-plane and in-plane potential gradients. It was thus demonstrated that the in-plane and out-of-plane asymmetry present in the Bi/GaAs(110) interface system plays an important role in the giant spin splitting. Our findings are anticipated to stimulate current experimental and theoretical studies for exploration of the giant spin splitting in other 1D electron systems formed on heterogeneous III-V semiconductor surfaces.

\vspace{0.4cm}

We thank Hyun Jung for his help in the initial stages of calculations. This work was supported by National Research Foundation of Korea (NRF) grant funded by the Korean Government (2015R1A2A2A01003248). The calculations were performed by KISTI supercomputing center through the strategic support program (KSC-2014-C3-011) for the supercomputing application research.

\noindent $^{*}$ Corresponding author: chojh@hanyang.ac.kr



\begin{thebibliography}{99}
\bibitem{lock} For the lower subband of the $SS$ state, the SAM is directed antiparallel to the OAM.
\end{thebibliography}
\end{document}